\documentclass[runningheads]{llncs}
\usepackage[T1]{fontenc}
\usepackage{graphicx}
\usepackage{multirow}
\usepackage{listings}

\begin{document}

\title{Adaptation of AI-accelerated CFD \\ simulations to the IPU platform}

\author{Paweł Rościszewski\inst{1} \and
Adam Krzywaniak\inst{1} \and
Sergio Iserte\inst{2,3} \and
Krzysztof Rojek\inst{4} \and
Paweł Gepner\inst{1,5}}

\authorrunning{P. Rościszewski et al.}

\institute{Graphcore Poland \\
\email{\{royr,adamk,pawelg\}@graphcore.ai} \and
Dpt. of Construction and Mechanical Engineering, Universitat Jaume I, Spain\\
\email{siserte@uji.es} \and
Barcelona Supercomputing Center, Spain \\ \email{sergio.iserte@bsc.es} \and
Institute of Computer and Information Sciences, Częstochowa University of Technology,
Częstochowa, Poland\\
\email{krojek@icis.pcz.pl} \and
Faculty of Mechanical and Industrial Engineering, Warsaw University of Technology,
Warszawa, Poland\\
\email{pawel.gepner@pw.edu.pl}}

\maketitle

\begin{abstract}
Intelligence Processing Units (IPU) have proven useful for many AI applications.
In this paper, we evaluate them within the emerging field of \emph{AI for simulation}, where traditional numerical simulations are supported by artificial intelligence approaches.
We focus specifically on a program for training machine learning models supporting a \emph{computational fluid dynamics} application. We use custom TensorFlow provided by the Poplar SDK to adapt the program for the IPU-POD16 platform and investigate its ease of use and performance scalability.
Training a model on data from OpenFOAM simulations allows us to get accurate simulation state predictions in test time.
We show how to utilize the \emph{popdist} library to overcome a performance bottleneck in feeding training data to the IPU on the host side, achieving up to 34\% speedup.
Due to communication overheads, using data parallelism to utilize two IPUs instead of one does not improve the throughput. However, once the intra-IPU costs have been paid, the hardware capabilities for inter-IPU communication allow for good scalability. Increasing the number of IPUs from 2 to 16 improves the throughput from 560.8 to 2805.8 samples/s.
\keywords{Intelligence processing unit \and Computational fluid dynamics \and Machine learning}
\end{abstract}

\section{Introduction}

One of the emerging trends in high performance computing (HPC) is supporting traditional numerical simulations with artificial intelligence (AI) approaches. While various names have been proposed for this field of research, such as \emph{simulation intelligence} \cite{lavin_simulation_2021} or \emph{cognitive simulation} \cite{wyatt_ii_is_2021}, we refer to it as simply as \emph{AI for simulation} \cite{ROJ2021,ROJ2021a}. Another significant trend is designing hardware architectures specifically for the type of workloads that are the backbone of AI \cite{survey,hardware_challenges,rosciszewski_impact_2019}. In this paper, we look into an \emph{AI for simulation} approach, where a machine learning (ML) model supports a computational fluid dynamics (CFD) application, and investigate how it can benefit from a AI-specific hardware architecture: the intelligence processing unit (IPU) processor.

The IPU is a computing accelerator specifically designed for machine learning computation. Each IPU has 1472 cores, with its own on-chip 624KiB SRAM memory per core. The combination of the core and the associated on-chip memory is named a tile. The tile Instruction Set Architecture (ISA) \cite{gepner} includes focused hardware elements such as Accumulating Matrix Product (AMP) and Slim Convolution (SLIC) units which allow to complete up to 64 multiply-add instructions per clock cycle. There are also hardware support instructions for random number generation and selected transcendental operations generally used in machine learning.
%The IPU supports 32-bit single-precision floating point FP32-IEEE, as well as FP16-IEEE 16-bit half-precision floating point numbers of data format with hardware stochastic rounding support.
Every tile runs 6 hardware execution threads in a time-sliced round-robin schedule, allowing instruction and memory latency to be hidden. With this mechanism, most instructions, including memory access and vectorised floating-point operations, complete within one thread cycle (6 clock cycles). Every thread represents a truly independent program. There is no restriction that threads run in groups executing the same program in lockstep, and no requirement that memory accesses are coalesced to achieve high SRAM bandwidth \cite{gepner}. 

IPU accelerators have proven useful for many AI applications, but employing them in \emph{AI for simulation} is a new area of research.
In this paper we adapt a training program for AI-accelerated CFD simulations to the IPU-based POD16 platform. This allows us to evaluate the models trained on the IPU-POD16 platform for the selected problem and investigate performance scalability of the training workload.
The remainder of the paper is organized as follows: references to related work are given in Section \ref{sec:related}, implementation details are described in Section \ref{sec:implementation}, experimental results are reported and discussed in Section \ref{sec:results}, while a summary is provided along with proposed future work directions in Section \ref{sec:summary}.

\section{Related work}\label{sec:related}

% We have divided the related work into two areas: augmenting simulations with AI (described in Section \ref{sec:simulations}) and utilizing various hardware for accelerating AI workloads (described in Section \ref{sec:hardware}).

% \subsection{AI-accelerated simulations}
% \label{sec:simulations}

Kochkov et al. in \cite{doi:10.1073/pnas.2101784118} summarized the applications of ML to accelerating numerical simulations and proposed the following classification:
\begin{itemize}
    \item supporting simulations with ML for better accuracy but no performance improvement;
    \item pure ML replacing the entire simulation, allowing for significant performance gains but weak on generalization (when new physical constraints are applied to previously trained model);
    \item hybrid approach replacing/accelerating iterative solvers inside the simulation without accuracy reduction.
\end{itemize}
We reviewed several papers which support such a classification of AI-accelerated simulations.

Maulik et. al in their work \cite{maulik_san_rasheed_vedula_2019} presented the results of two-dimensional Kraichan turbulence subgrid modeling with a novel data-driven neural network support for predicting the turbulence source. Their work aimed to improve the accuracy of modeling without focusing on increasing its performance.

Kim et al. in their paper \cite{doi.org/10.1111/cgf.13619} proposed a generative model called DeepFluids to synthesize fluid simulations from a set of reduced parameters. They train a convolutional neural network (CNN) for predicting the fluid velocity fields. In their work they propose a fluid-specific loss function to improve the convergence of the trained model. The aim of their work is to replace the simulation in order to use the trained ML model in inference mode and improve the performance of velocity fields reconstruction up to 700x.

Wiewel et al. in their work \cite{doi.org/10.1111/cgf.13620} proposed an approach based on the long short-term memory (LSTM) network for fluid flow modeling, i.e. to predict the changes of pressure fields over time. They achieved practical speed-ups with neural network-based simulation of 3D+time functions of a physics system.

Ribeiro et al. in their paper \cite{ribeiro_deepcfd_2021} presented  a CNN-based model called DeepCFD, that efficiently approximates solutions for the problem of non-uniform steady laminar flows. Their proposed model is able to learn complete solutions of the Navier-Stokes equations, for both velocity and pressure fields, directly from ground-truth data generated using a state-of-the-art CFD code. The predictions of the proposed model allow for achieving up to 1000x speedup in obtaining the resulting velocity and pressure fields, when comparing classical simulation on CPU with CNN model running on GPU.

Thuerey et. al in their work \cite{deep-flow-prediction} investigated the accuracy of deep learning models for the inference of Reynolds-Averaged Navier-Stokes solutions. Their best results allowed them to obtain mean relative pressure and velocity error of less than 3\% across a range of previously unseen airfoil shapes.

Um et al. in their paper \cite{10.5555/3495724.3496237} present a hybrid approach called Solver-in-the-loop. By integrating the learned function into a differentiable physics pipeline, the corrections can interact with the physical system, alter the states, and receive gradients about the future performance of these modifications. This provided the model with realistic input distributions that take previous corrections into account, yielding improvements in accuracy with stable rollouts of several hundred recurrent evaluation steps and surpassing even tailored supervised variants.

In this paper we evaluate the LSTM-based approach for predicting the fluid flow in a homogenization tank which aims to replace the simulation with an OpenFOAM numerical solver. We evaluate the LSTM model on IPU, a new AI-dedicated massively parallel hardware accelerator.

\section{Implementation}\label{sec:implementation}

In this section we describe the details behind the proposed implementation. First, in Section \ref{sec:model} we describe the original implementation of the model selected for adaptation. Section \ref{sec:pod} contains a detailed description of the hardware configuration used for the experiments. The basic process of porting the training application to the IPU platform is described in Section \ref{sec:porting}. Additionally, Section \ref{sec:popdist}, describes the improvements that we introduced using the popdist library to alleviate data loader limitations.

\subsection{The original Model for accelerating CFD simulations}\label{sec:model}

\begin{figure}[htp]
    \centering
    \includegraphics[trim={0 0 0 0}, clip, width=0.6\linewidth]{"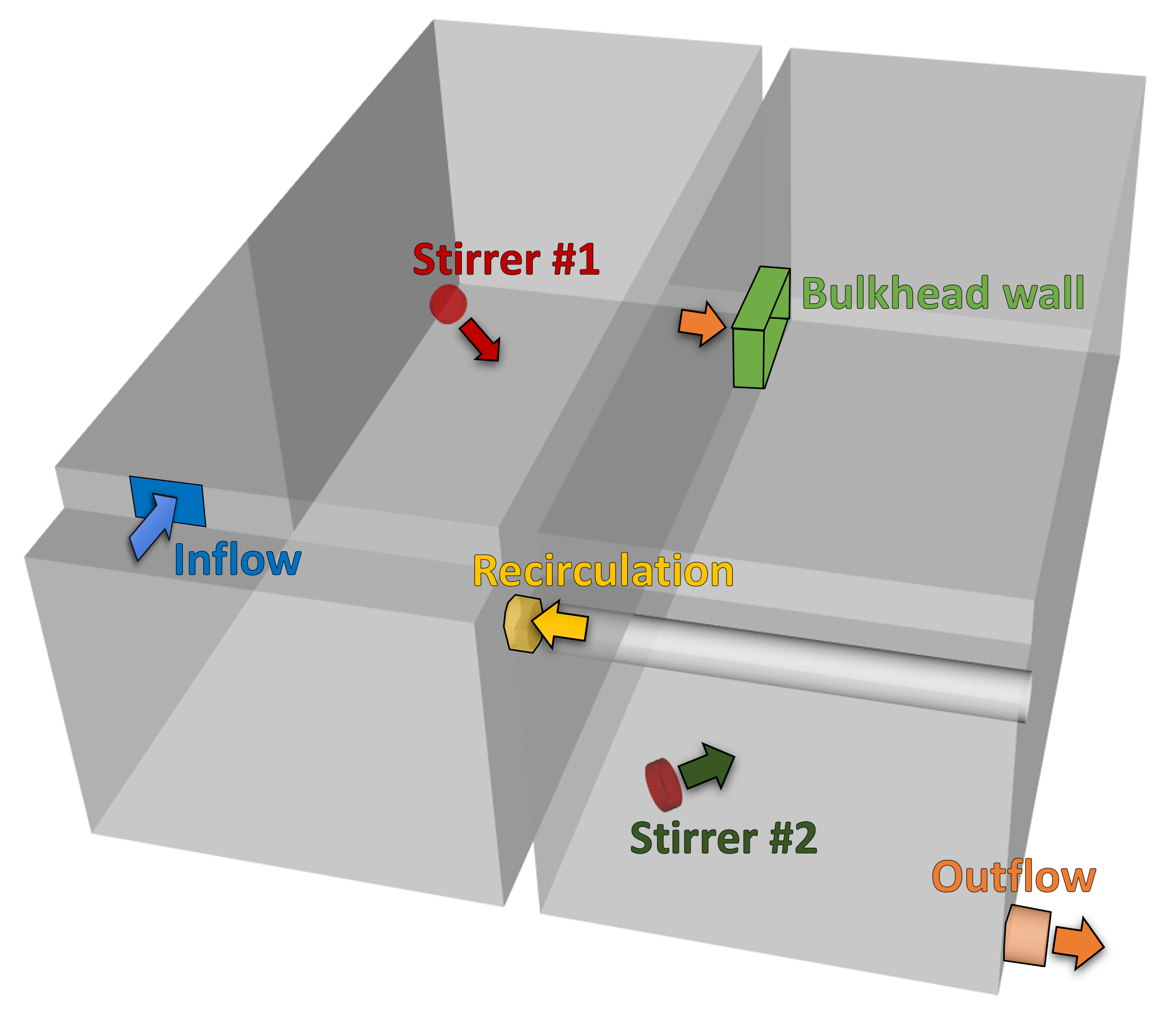"}
    \caption{Geometry of the reactor under study~\cite{Iserte2021}. Arrows represent the flow direction on the highlighted areas.}
    \label{fig:tank}
\end{figure}

The case study selected for this paper trains a ML model for accelerating CFD simulations of an industrial homogenization tank.
The tank is composed of two interconnected subtanks of $10 m$ length, $5 m$ width, and $5 m$ depth each.
Figure~\ref{fig:tank} depicts the geometry of the tank.
The figure highlights the location of the areas of interest. They do not correspond to regular walls and can be parametrized.
The flow enters the tank through \textit{Inflow}. 
The flow is driven to \textit{Outflow} through \textit{Bulkhead wall}. 
Part of the flow is fed back from the second subtank to the first one using \textit{Recirculator}.
Finally, stirrers inside the tank (\textit{Stirrer \#1 and \#2}) are responsible for impelling the flow.

We have simulated 131 different configurations of the case under study with OpenFOAM\footnote{\url{http://www.openfoam.com}}.
In these simulations, the values of in \textit{Inlet} and \textit{Recirculation} are varied within a minimum and a maximum limit.
The OpenFOAM solver models a transient incompressible flow using the Unsteady Reynolds-averaged Navier-Stokes (URANS) equations.
The state of the simulation is adjusted to a write interval of 10 seconds of simulated time. 
The flow is evolved until the second 4,201, which is translated into 420 stored states per executed simulation.
In this regard, taking it all into account, the number of simulated cases, states per case, the cells in the domain, and the velocity dimensions, generate an eventual dataset with a shape $131 \times 420 \times 125,565 \times 3$.

Before feeding the trainable model, each velocity dimension is normalized to have a distribution of mean zero and a standard deviation of one. 
Moreover, the dataset is split into train and test subsets. 
For this purpose, cases are shuffled and 80\% of them (104 cases) are assigned to the training dataset, while the remaining (27 cases) are assigned to the testing dataset. 
Notice that 20\% of training cases (20 cases) are used for cross-validating the learning.

In order to capture the temporal dependencies in the data, a sequence to sequence model is trained, where features representing 3 consecutive simulation states are used as the input sequence, while output represents 1 succeeding simulation state. The training program is implemented in Python using TensorFlow and the model is constructed sequentially using the Keras API, as shown in Listing \ref{lst:model_impl}. The main building blocks of the model are the encoder and decoder LSTM layers with 10 hidden units each. Repeated copies of the encoder output are used as the input for the decoder. Finally, temporal slices of the decoder output are used by two dense layers. 
The rectified linear unit (ReLU) function is used for activation and Adam optimizer is used for training.

It should be noted that considering the input sequence length, the number of cells in the domain and the velocity dimensions, the input sequence dimensionality is $3 \times 125,565 \times 3$ giving $1,130,085$ features per sample. This, in combination with relatively small hidden state numbers in the model layers, makes the training workload highly I/O-bound.

\begin{lstlisting}[caption=Implementation of the model layers, label=lst:model_impl, basicstyle=\scriptsize]
from tensorflow.keras.layers import LSTM, RepeatVector,
                                    Dense, TimeDistributed
...
model = Sequential()
model.add(LSTM(10, activation='relu',
               input_shape=(n_timesteps, n_features), 
               return_sequences=False))

model.add(RepeatVector(n_outputs))
model.add(LSTM(10, activation='relu', return_sequences=True))

model.add(TimeDistributed(Dense(10, activation='relu')))
model.add(TimeDistributed(Dense(n_features)))

opt = tf.keras.optimizers.Adam(learning_rate=0.00025)

model.compile(loss='mae', optimizer=opt)

model.fit(train, epochs=n_epochs, verbose=1, shuffle=True,
          steps_per_epoch=steps_per_epoch, callbacks=callbacks)
\end{lstlisting}

\subsection{The IPU processor, IPU-M2000 system and IPU-POD16 configuration}
\label{sec:pod}

From the hardware definition IPUs are distributed memory, massively parallel, multiple-instruction multiple-data (MIMD) devices. With 1472 tiles, the IPU has just under 900 MB of memory in total. This local memory is the only memory directly accessible by tile instructions. It is used for both the code and the data used by that tile. There is no shared memory access between tiles.
%The tile uses a contiguous unsigned 21-bit address space, beginning at address 0x0. The effect of accessing unpopulated memory addresses is undefined. Memory parity errors can occur when data is read from memory, for example, by a load instruction or an instruction fetch. A parity error detected in a fetched instruction prevents the execution of that instruction.
Tiles cannot directly access each others’ memory but can communicate via message passing using an all-to-all high bandwidth exchange (theoretical 8 TB/s). The memory has very low-latency (6 cycles) and ultra-high bandwidth (theoretical 47.5 TB/s). The whole chip is built on the budget of 59.4 billion transistors using the TSMC 7nm manufacturing process \cite{gepner}. 

\begin{figure}[tbp]
\centering
\includegraphics[width=0.75\hsize]{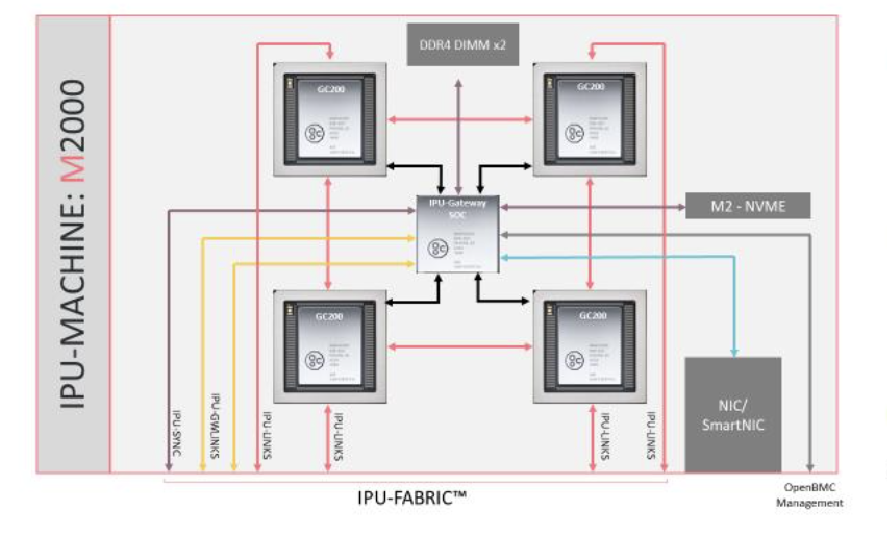}
\caption{Schematic and building block of IPU-M2000 Machine \cite{gepner}}
\label{fig:m2000}
\end{figure}

The Graphcore IPU-M2000 system is essentially a 1U server utilizing 4 IPUs. It includes also a gateway chip which connects IPUs into the compute domain and provides access to the DRAM, two 100Gbps IPU-Fabric Links, a PCIe slot for standard Smart NICs, two 1GbE Open BMC management interfaces, and access to an M.2 slot. Figure \ref{fig:m2000} shows the block diagram of the IPU-M2000 system. The host system accesses the IPU-M2000 platform over 100Gb Ethernet with ROCE (RDMA over Converged Ethernet) with very low-latency access. Such an implementation based on Ethernet avoids the bottlenecks and costs of PCIe connectors and PCIe switches. This enables a flexible host CPU to accelerators combination and provides scaling from a single IPU-M2000 system to massive supercomputer scale including 64,000 IPUs, all networked over standard networking at a lower cost and providing much more flexibility than using e.g., InfiniBand \cite{Freud}.

\begin{figure}[tbp]
\centering
\includegraphics[width=0.8\hsize]{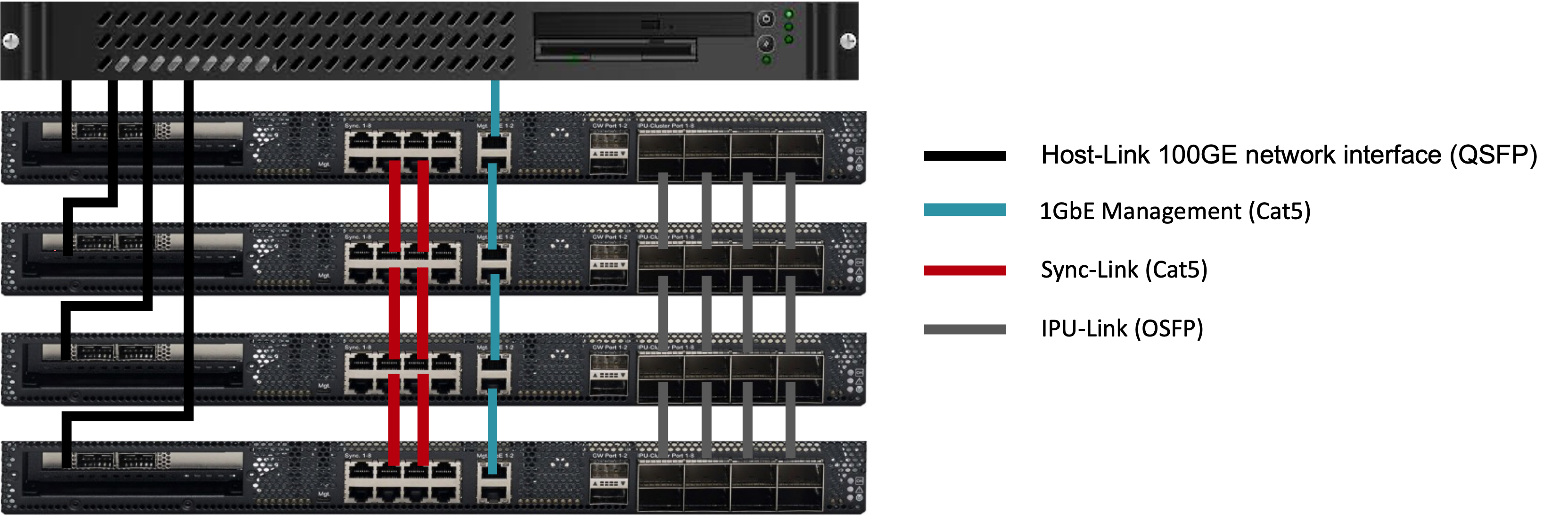}
\caption{IPU-POD16 direct attach configuration \cite{gepner}}
\label{fig:pod16}
\end{figure}

IPU-Fabric is a totally new scale-out fabric designed from the ground up to support the needs of machine intelligence communication. IPU–Fabric is natively integrated into the IPU processors and IPU-M2000 system. A key difference between IPU-Fabric and other proprietary fabrics is the usage of Compiled Communication and Bulk Synchronous Parallel protocol; both these elements provide deterministic communication behaviour. Every IPU has dedicated IPU-Links providing 64GB/s of bidirectional bandwidth and an aggregate bandwidth per chip of 320 GB/s. Each IPU-M2000 has 8 external IPU-Links for intra-rack scale out using OSFP copper cables. The intra-rack configuration called IPU-POD16 contains 4 IPU-M2000s connected into a single instance with a daisy chain topology utilizing IPU-Links. Host-Link connectivity is provided from the Gateway through a PCIe NIC or SmartNIC card. Figure \ref{fig:pod16} shows the IPU-POD16 configuration \cite{Freud}.

The memory model for the IPU-Machine is also quite unique. In addition to in-IPU Memory, each IPU-M2000 system has DDR memory available to the four IPUs. This DDR memory is used differently from memory found in CPUs or GPUs. Instead of a memory hierarchy that requires swapping data and code from the host memory store to the accelerator’s memory, the Poplar Graph Compiler creates deterministic code-memory relationships in both the memory on the IPU tile and the DDR memory. In fact, the IPU-M2000 system can use this additional memory in stand-alone mode for inference processing without any attachment to a host server. And thanks to the bulk synchronous parallel (BSP) model compiling both computation and communication, the network communication overhead is kept to a minimum compared to traditional messaging or shared memory constructs commonly used for parallel processing.\\ 
Built-in fabrics are becoming a necessity for AI accelerators since model sizes are increasing dramatically, some containing billions of parameters. These large models must be distributed across hundreds or thousands of processors to solve problems in a reasonable time. Graphcore’s hybrid model uses a proprietary IPU-Link fabric to communicate across the tiles in an IPU and adjacent rack IPUs, while tunnelling the IPU-Link protocol across standard 100GbE for rack-to-rack scale-out supporting larger configurations \cite{Freud}.
%This disaggregated scaling model is the most important feature of IPU-M2000 based systems and, together with IPU- Fabric, enables a flexible disaggregation model, allowing the user to configure multiple accelerators on the fly without constraints from a predetermined scenario. \\

\subsection{Porting the training program to the IPU platform}\label{sec:porting}

The IPU is based on a sophisticated architecture which offers, to our knowledge, the first ever BSP model implementation in hardware. Fortunately, users do not have to be parallel processing experts to benefit from the performance opportunities offered by the IPU accelerator. The hardware comes with a comprehensive software ecosystem \footnote{\url{https://www.graphcore.ai/developer}} with the Poplar SDK \footnote{\url{https://docs.graphcore.ai/projects/sdk-overview/en/latest/index.html}}, a complete tool chain that enables the user to exploit IPU features. The SDK includes a graph compiler responsible for handling the scheduling and work partitioning of large parallel programs including memory control.
To provide maximum possible ease of use, the SDK is integrated with a number of industry-standard ML frameworks. In this paper, we benefit from such an integration with TensorFlow, which requires the user to employ \emph{pip} to install a separate \emph{tensorflow} package provided with the Poplar SDK. Using this approach, porting the original code to the IPU platform requires only a few changes, as outlined in Listing \ref{lst:ipu_impl}.

\begin{lstlisting}[caption=Code changes required to port the program to the IPU, label=lst:ipu_impl, basicstyle=\scriptsize]
from tensorflow.python import ipu

config = ipu.config.IPUConfig()
config.auto_select_ipus = FLAGS.num_replicas
config.configure_ipu_system()

strategy = ipu.ipu_strategy.IPUStrategy()

with strategy.scope():
    <code from Listing 1.1>
\end{lstlisting}

Running the code on the IPU requires the user to import the corresponding module \emph{tensorflow.python.ipu}  and use it to configure the IPU system as well as place the adequate variables on the IPU. Running the training on multiple IPUs using data parallelism is as simple as setting the \emph{auto\_select\_ipus} configuration parameter to the desired value. Tensor and operation placement is performed by wrapping the original code in the scope of a custom implementation of a TensorFlow distribution strategy. Additionally, to avoid frequent host-IPU synchronization, it is worth setting the \emph{steps\_per\_execution} parameter of the \emph{model.compile()} function to a large value. We use the number of steps per epoch as a rule of thumb in order to run the whole epoch on the IPU before returning to the host.
This straightforward approach to porting the code to IPU benefits from the ease of use of the Poplar SDK's TensorFlow integration. 

\subsection{Using the popdist library to remove the I/O bottleneck}\label{sec:popdist}

In many cases a simple porting procedure described in Section \ref{sec:porting} would be sufficient for optimal utilization of the IPU platform. However, as shown in Section \ref{sec:results}, in the case of the investigated IO-bound CFD application, using a single Python process for feeding multiple IPUs with input data results in a I/O bottleneck.

To remove this bottleneck, we used the \emph{poprun} tool associated with the Poplar distributed configuration library (\emph{popdist})\footnote{https://docs.graphcore.ai/projects/poprun-user-guide/en/latest/configuration.html} to execute a separate system process per each IPU. The crucial code changes required are shown in Listing \ref{lst:ipu_popdist}.

\begin{lstlisting}[caption=Code changes required to run the training in a distributed setup, label=lst:ipu_popdist, basicstyle=\scriptsize]
import popdist
from tensorflow.python.ipu import horovod as hvd
from tensorflow.python.ipu.horovod import popdist_strategy
...
popdist.tensorflow.set_ipu_config(config, ipus_per_replica=1)
hvd.init()
...
train = train.shard(num_shards=popdist.getNumInstances(),
                    index=popdist.getInstanceIndex())
...
strategy = popdist_strategy.PopDistStrategy()
\end{lstlisting}

\emph{Popdist} allows the user to automatically configure the desired number of IPUs per model replica. It is used along with an implementation of the Horovod communication scheme \cite{sergeev_horovod:_2018}. Shards of the training dataset are selected accordingly to the number of instances executed by the \emph{poprun} tool and the corresponding process instance numbers. Finally, instead of the standard IPUStrategy, the PopDistStrategy class ensures the proper variable placement in the context of distributed execution.

\section{Experimental results}\label{sec:results}

In order to evaluate the usefulness of the IPU-POD16 platform for the application we are focusing on in this paper, first we utilized it to train a model and investigated its accuracy. The findings are described in Section \ref{sec:verification}. Then, to assess its performance capabilities, we measured training throughput depending on the number of used IPUs and chosen implementation.
%Additionally, since we noticed that the half-float (FP16) precision is good enough for storing input data, we investigated the influence of limiting data representation precision from full float (FP32) to FP16 on the performance of the IPU system. It should be noted that precision was limited only for the input features, while the model weights were always stored in FP32.
The performance results are provided in Section \ref{sec:scalability}. The experiments were run on an IPU-POD16 with 16 IPU-M2000 IPU chips using Graphcore TensorFlow-2.6.3 and Keras 2.6.0 on top of Poplar SDK r2.6.0.

%TODO (SI): do you have a paper that can be used as reference that FP16 is enough precision for storing the input data?

\subsection{Model verification}\label{sec:verification}

%In order to verify the results, first the dataset must be described.

To develop a model for verification, we executed ten training sessions with a random selection of learning rate between 1e-7 and 1e-5. The runs were stopped when the validation loss has not improved more than 0.0001 for 10 epochs. Out of the ten trained models we selected the one that performed best on the validation set. The accuracy results for this model are presented in Table \ref{tab:model_accuracy}.

To estimate the accuracy, we used statistical metrics such as RMSE (root-mean-square-error) and correlation coefficients that measure the extent to which two variables tend to change together. These coefficients describe both the strength and the direction of the relationship. Here, we use two coefficients, including the Pearson correlation which estimates the linear relationship between two continuous variables, as well as the Spearman correlation which assesses the monotonic relationship between two continuous or ordinal variables. The correlation coefficients can return values from -1 to 1. The RMSE statistic shows that the error is below 0.08 for all the results. Since the range of data is from 0 to 1.1 we conclude that the differences are below 8\% of the maximum value for the 10th-time step and below 1\% for the steady-state. The correlation coefficients show a strong dependency between trends of the predicted and real values (>0.9 for all the time steps).

\begin{table}
\centering
\begin{tabular}{| c | c | c | c |} 
 \hline
 Step & Pearson's correlation & Spearman's correlation & Root mean squared error \\ [0.5ex] 
 \hline\hline
 10 & 0.949 & 0.917 & 0.078 \\ 
 \hline
  20 & 0.989 & 0.982 & 0.038 \\ 
 \hline
  100 & 1.000 & 0.999 & 0.008 \\ 
 \hline
\end{tabular}
\caption{Accuracy of the trained model for the selected steps of simulation.}
\label{tab:model_accuracy}
\end{table}

\subsection{Performance and scalability}\label{sec:scalability}
% TODO!!!!!!!!!!!!!!!: How to comment on the lack of comparison vs GPU/CPU? How to tell if the reported throughput is good?

%Figures \ref{fig:scalability_fp32} and \ref{fig:scalability_fp16} present throughput results for FP32 and FP16 input data types respectively, depending on the number of utilized IPUs and implementation variant.
Table \ref{tab:scalability_results} shows training throughput depending on the number of utilized IPUs and implementation variant averaged from the aforementioned five runs, additionally providing standard deviation.
The number of IPUs corresponds to the number of model replicas in the "data parallel" scheme used for training parallelization. The "single process" variant is described in Section \ref{sec:porting} while the "popdist" variant is described in Section \ref{sec:popdist}. We performed five runs for each parameter combination. In each run, we executed four training epochs and measured the throughput for the three last epochs as the total number of used samples divided by execution time. We treated the first epoch as a warm-up.
%Table \ref{tab:scalability_results} contains detailed results averaged from the aforementioned five runs, additionally providing standard deviation. 

While IPUs do not have a particularly high memory capacity, they do not require large batch sizes to achieve good performance, so for all experiments we used mini-batches containing one training sample. As training data, we used random samples generated on the host side, so that the benchmark measures the capability of the host + IPU system as a whole, without considering potential limitations of storage I/O overheads. To overcome the limitations of FP16 data handling on the CPU side, in two cases of single process implementation (8 and 16 IPUs) we used non-standard, increased buffer sizes in the internal TensorFlow data queue.

%\begin{figure}
%  \centering
%  \includegraphics[width=.8\linewidth]{images/scalability_fp32.png}
%  \caption{Throughput results for FP32 input data type, depending on the implementation variant and the number of utilized IPUs}
%  \label{fig:scalability_fp32}
%\end{figure}

%\begin{figure}
%  \centering
%  \includegraphics[width=.9\linewidth]{images/scalability_fp16.png}
%  \caption{Throughput results for FP16 input data type, depending on the implementation variant and the number of utilized IPUs}
%  \label{fig:scalability_fp16}
%\end{figure}

The results allow us to draw the following conclusions. Firstly, most of the results are statistically significant, with exceptions in the cases where 4, 8 and 16 IPUs are used by the single process implementation. The configuration that results in the most variable results (16 IPUs, single process) is also the one which benefits the most from switching to \emph{popdist} (34\% speedup). We performed detailed profiling of the program to determine that there is a  bottleneck on the host side: multi-threading limitations of Python result in slow data pre-processing and populating the input data queue by the CPU. As a result, the more IPUs that are used, the more likely that they are starved.

\begin{table}
    \centering
        \begin{tabular}{ || c || p{2.5cm} | p{2.5cm} || }
         \hline
         \multirow{2}{3cm}{No. of IPUs utilized} & \multicolumn{2}{|c||}{average throughput} \\ 
         \cline{2-3}
          & single process & popdist \\
         \hline
         1 & $571.8 \pm 4.31$ & $574.4 \pm 3.20$ \\
         2 & $558.8 \pm 3.92$ & $560.8 \pm 1.94$ \\
         4 & $862.8 \pm 7.14$ & $871.4 \pm 1.36$ \\
         8 & $1344.2 \pm 8.35$ & $1566.4 \pm 1.02$ \\
         16 & $2099.8 \pm 193.19$ & \textbf{2805.8} $\pm 1.17$ \\
         \hline
        \end{tabular}
    \caption{Throughput (samples/s) depending on implementation variant and number of utilized IPUs}
    \label{tab:scalability_results}
\end{table}

Another interesting observation is related to scalability: increasing the number of used IPUs from 1 to 2 doesn't significantly improve the throughput, and even makes it slightly worse. At the same time, increasing the number of IPUs 8-fold from 2 to 16 improves the throughput around 5-fold, which is relatively good scalability, considering the characteristics of data-parallel deep neural network training.
% TODO(PR) citation for data-parallel scalability
Again, the reason for the lack of scalability between 1-2 IPUs has been determined through detailed profiling. In this case, the bottleneck is on the IPU side: for this particular model, the overhead of introducing additional buffers and exchange operations makes the data-parallel implementation significantly slower on a single IPU. % TODO(PR): consider citing PopVision

%Finally, we notice that using the FP16 type for storing (and, more importantly, transmitting from host to the IPU) significantly and universally improves the system throughput. For the best-performing configuration of 16 IPUs utilized by the \emph{popdist} implementation, using FP16 instead of FP32 enables the throughput to be improved by 12\%. This result is expected, especially considering the IO-bound nature of the training program under investigation here. As a result, the best training throughput achieved on the IPU-POD16 is 2808.8 samples/s.

\section{Summary and future work}\label{sec:summary}

In this paper, we adopt a deep neural network training application from the \emph{AI for simulation} field for the IPU platform, demonstrating the ease of use provided by the Poplar SDK software ecosystem. Training a model on data from traditional CFD simulations allows us to get accurate simulation state predictions in test time. Investigating the performance of the training on the IPU-POD16 platform reveals that the main bottleneck of this particular application is feeding training data to the IPU on the host side. We show how to utilize the \emph{popdist} library to overcome the limitations of host-side data loading.
Scaling of the program is limited to a small scale of 1-2 IPUs by communication overheads.
However, once the intra-IPU costs have been paid, the hardware capabilities for inter-IPU communication allow for good scalability. 

%- Casting the input data to the FP16 data type significantly improves the training performance - data representation should be always considered.

In the future, we would like to investigate the scalability of the IPU platform further, utilizing a larger platform such as the IPU-POD64. It could be also beneficial to use the FP8 data type to increase training performance.
The predictive model introduced in this work can be leveraged in hybrid CFD-DL solvers such as that presented in~\cite{Iserte22}. 
This solver alternates stages of CFD simulation with predictions made by a DL engine in order to reduce the time-to-solution. In their paper, the authors are able to accelerate the simulation interleaving predictions during the CFD simulation. That module could be easily substituted by the IPU-trained model for inference.

\section*{Acknowledgements}
The authors would like to thank Grzegorz Andrejczuk for his ideas and help with investigating data loading overheads. Big thanks to Charis Fisher for her support and valuable comments.
Researcher Sergio Iserte was supported by the postdoctoral fellowship APOSTD/2020/026 from Valencian Region Government (GVA) and European Social Funds (ESF).
CFD Simulations were executed on Tirant III cluster of the \textit{Servei d'Informàtica} of the University of Valencia (UV).

\bibliographystyle{splncs04}
\bibliography{ppam22}

\end{document}